\documentclass[showpacs,reprint,preprintnumbers,amsmath,amssymb,aip,apl]{revtex4-1}

\usepackage[sort&compress]{natbib}
\usepackage{graphicx}
\usepackage{dcolumn}
\usepackage{bm}
\usepackage{amsmath}

\newcommand{\oC}{$^\mathrm{o}$C}

\newcommand{\heusler}{Co$_2$FeAl$_{0.4}$Si$_{0.6}$}

\begin{document}

\title{Spin configurations in Co$_2$FeAl$_{0.4}$Si$_{0.6}$ Heusler alloy thin film elements}

\author{C. A. F. Vaz}
\email[Corresponding author. Email: ]{carlos.vaz@cantab.net}%
\affiliation{SwissFEL, Paul Scherrer Institut, 5232 Villigen PSI, Switzerland}

\author{J. Rhensius}
\affiliation{SwissFEL, Paul Scherrer Institut, 5232 Villigen PSI, Switzerland}
\affiliation{Laboratory for Nanomagnetism and Spin Dynamics, Ecole Polytechnique F\'{e}d\'{e}rale de Lausanne (EPFL), 1015 Lausanne, Switzerland}
\affiliation{Laboratory for Micro- and Nanotechnology, Paul Scherrer Institut, 5232 Villigen PSI, Switzerland}
\affiliation{Fachbereich Physik, Universit\"{a}t Konstanz, Universit\"{a}tsstra\ss e 10, 78457 Konstanz, Germany}

\author{J. Heidler}
\affiliation{SwissFEL, Paul Scherrer Institut, 5232 Villigen PSI, Switzerland}
\affiliation{Laboratory for Nanomagnetism and Spin Dynamics, Ecole Polytechnique F\'{e}d\'{e}rale de Lausanne (EPFL), 1015 Lausanne, Switzerland}

\author{P. Wohlh{\"u}ter}
\affiliation{SwissFEL, Paul Scherrer Institut, 5232 Villigen PSI, Switzerland}
\affiliation{Laboratory for Nanomagnetism and Spin Dynamics, Ecole Polytechnique F\'{e}d\'{e}rale de Lausanne (EPFL), 1015 Lausanne, Switzerland}
\affiliation{Fachbereich Physik, Universit\"{a}t Konstanz, Universit\"{a}tsstra\ss e 10, 78457 Konstanz, Germany}

\author{A. Bisig}
\affiliation{SwissFEL, Paul Scherrer Institut, 5232 Villigen PSI, Switzerland}
\affiliation{Laboratory for Nanomagnetism and Spin Dynamics, Ecole Polytechnique F\'{e}d\'{e}rale de Lausanne (EPFL), 1015 Lausanne, Switzerland}
\affiliation{Fachbereich Physik, Universit\"{a}t Konstanz, Universit\"{a}tsstra\ss e 10, 78457 Konstanz, Germany}

\author{H. S. K{\"o}rner}
\affiliation{SwissFEL, Paul Scherrer Institut, 5232 Villigen PSI, Switzerland}
\affiliation{Laboratory for Nanomagnetism and Spin Dynamics, Ecole Polytechnique F\'{e}d\'{e}rale de Lausanne (EPFL), 1015 Lausanne, Switzerland}
\affiliation{Fachbereich Physik, Universit\"{a}t Konstanz, Universit\"{a}tsstra\ss e 10, 78457 Konstanz, Germany}

\author{T. O. Mentes}
\affiliation{Sincrotrone Trieste, 34149 Basovizza-Trieste, Italy}

\author{A.~Locatelli}
\affiliation{Sincrotrone Trieste, 34149 Basovizza-Trieste, Italy}

\author{L. Le Guyader}
\affiliation{Swiss Light Source, Paul Scherrer Institut, 5232 Villigen, Switzerland}

\author{F. Nolting}
\affiliation{Swiss Light Source, Paul Scherrer Institut, 5232 Villigen, Switzerland}

\author{T. Graf}
\affiliation{Institute for Analytical and Inorganic Chemistry, Johannes Gutenberg-Universt\"{a}t, 55099 Mainz, Germany}

\author{C. Felser}
\affiliation{Institute for Analytical and Inorganic Chemistry, Johannes Gutenberg-Universt\"{a}t, 55099 Mainz, Germany}

\author{L. J. Heyderman}
\affiliation{Laboratory for Micro- and Nanotechnology, Paul Scherrer Institut, 5232 Villigen PSI, Switzerland}

\author{M. Kl\"{a}ui}
\email[Current address: ]{Institut f\"ur Physik, Johannes Gutenberg-Universt\"{a}t, 55099 Mainz, Germany}%
\affiliation{SwissFEL, Paul Scherrer Institut, 5232 Villigen PSI, Switzerland}
\affiliation{Laboratory for Nanomagnetism and Spin Dynamics, Ecole Polytechnique F\'{e}d\'{e}rale de Lausanne (EPFL), 1015 Lausanne, Switzerland}
\affiliation{Fachbereich Physik, Universit\"{a}t Konstanz, Universit\"{a}tsstra\ss e 10, 78457 Konstanz, Germany}

\date{\today}

\begin{abstract}
We determine experimentally the spin structure of half-metallic Co$_2$FeAl$_{0.4}$Si$_{0.6}$ Heusler alloy elements using magnetic microscopy. Following magnetic saturation, the dominant magnetic states consist of quasi-uniform configurations, where a strong influence from the magnetocrystalline anisotropy is visible. Heating experiments show the stability of the spin configuration of domain walls in confined geometries up to 800 K. The switching temperature for the transition from transverse to vortex walls in ring elements is found to increase with ring width, an effect attributed to structural changes and consequent changes in magnetic anisotropy, which start to occur in the narrower elements at lower temperatures.
\end{abstract}



\maketitle

Heusler alloys are materials systems characterized by rich electronic and magnetic properties, such as half-metallic behavior, large magneto-optic constants, and shape memory.\cite{GFP11,BEJ83} The combination of high magnetic critical temperatures, large magnetization, and half-metallicity, makes the Heusler alloys particularly interesting candidates for spintronics devices, for example, as spin injectors in spin polarized field effect transistors (spin-FET), magnetic layers in tunnel junctions, or as spin polarized materials for spintronics applications,\cite{GFP11,TGHH10,Takanashi10,IVT+05a} while recently a magnetoelectric coupling\cite{VHAR10} has been predicted in a Heusler/ferroelectric junction.\cite{YSP07} In half-metals, one of the spin sub-bands is semiconducting, rendering electron transport fully spin-polarized; for a robust half-metallic behavior, a large separation between the Fermi energy and the edges of the semiconducting sub-band is desirable, to avoid thermal excitation and spin-flip scattering of electrons to the conduction sub-band. Such tuning of the Fermi level position is possible by adjusting the doping level $x$ in half-metallic Co$_2$FeAl$_{1-x}$Si$_{x}$, where it is found that for $x\sim 0.5$, the Fermi Level lies in the middle of the semiconducting spin sub-band.\cite{FF07,NRG+07} Another advantage of this composition is that the system becomes less sensitive to atomic disorder, whose effect is to introduce edge states at the band gap of the semiconducting sub-band and to reduce the spin polarization.\cite{FF07,NRG+07} These factors explain the large tunneling magnetoresistance (up to 160\%)\cite{TIM+06,SSW+09} and the efficient spin-transfer switching in spin-valve nanopillar structures using this particular composition.\cite{SKF+10}

To utilize these unique features, a detailed knowledge of the spin structure in these materials at the nanoscale is necessary. In this work, we study the magnetic spin configuration of patterned Heusler alloy \heusler\ thin films using x-ray magnetic circular dichroism photoemission electron microscopy (XMCD-PEEM). We find that the spin configurations in small elements (below 2~$\mu$m) are quasi-uniform, while well defined domain walls are formed in narrow structures. The magnetic configurations are found to be stable up to 800 K, making it particularly suitable for room temperature device applications.

The structures investigated in this study consist of arrays of discs, rings, squares, rectangles, triangles, and zig-zag wires of various sizes, fabricated by d.c.\ magnetron sputtering of [1.5 nm] Ru/ [$t$] \heusler\ films, where $t =15, 30$ nm, on a polymethyl methacrylate (PMMA) resist spincoated on a Cr [10 nm]/MgO(001), defined using e-beam lithography, and subsequent lift-off. The films were deposited at room temperature and the samples were afterwards annealed at 500\oC\ in a nitrogen atmosphere for 30 min. X-ray diffraction measurements on control samples grown under the same conditions indicate that the as-deposited \heusler\ films grow epitaxially on Cr/MgO(001) with B2-type order, while after annealing L21-type order was observed.\cite{GCG+11} The continuous Cr layer prevents sample charging during magnetic imaging using XMCD-PEEM. In this technique, a photoemission electron microscope is used to image the local difference in light absorption at resonance for left and right circularly polarized light. The magnetic contrast depends on the angle between the helicity vector and magnetization, being maximum when they are parallel or anti-parallel. For the XMCD measurements presented here, the x-ray light was tuned to the Co $L$ edge, which displays larger intensity than the Fe $L$ edge for \heusler. Due to the grazing incidence of the photon beam, our measurements are mainly sensitive to the in-plane component of the magnetization. Magnetic force microscopy imaging of selected elements (not shown) confirms the  in-plane orientation of the magnetization.  2D  micromagnetic simulations were carried out with the Object Oriented MicroMagnetic Framework (OOMMF)\cite{OOMMF} package using the materials parameters: $M_s = 1000$ emu/cm$^3$, $A = 2.3 \times 10^{-6}$ erg/cm, $K_1 = -9\times 10^4$ erg/cm$^3$, and 5 nm in-plane cell size.\cite{TGHH10,GCG+11}

\begin{figure}[tbh]
\centering
\includegraphics[width=8.5cm]{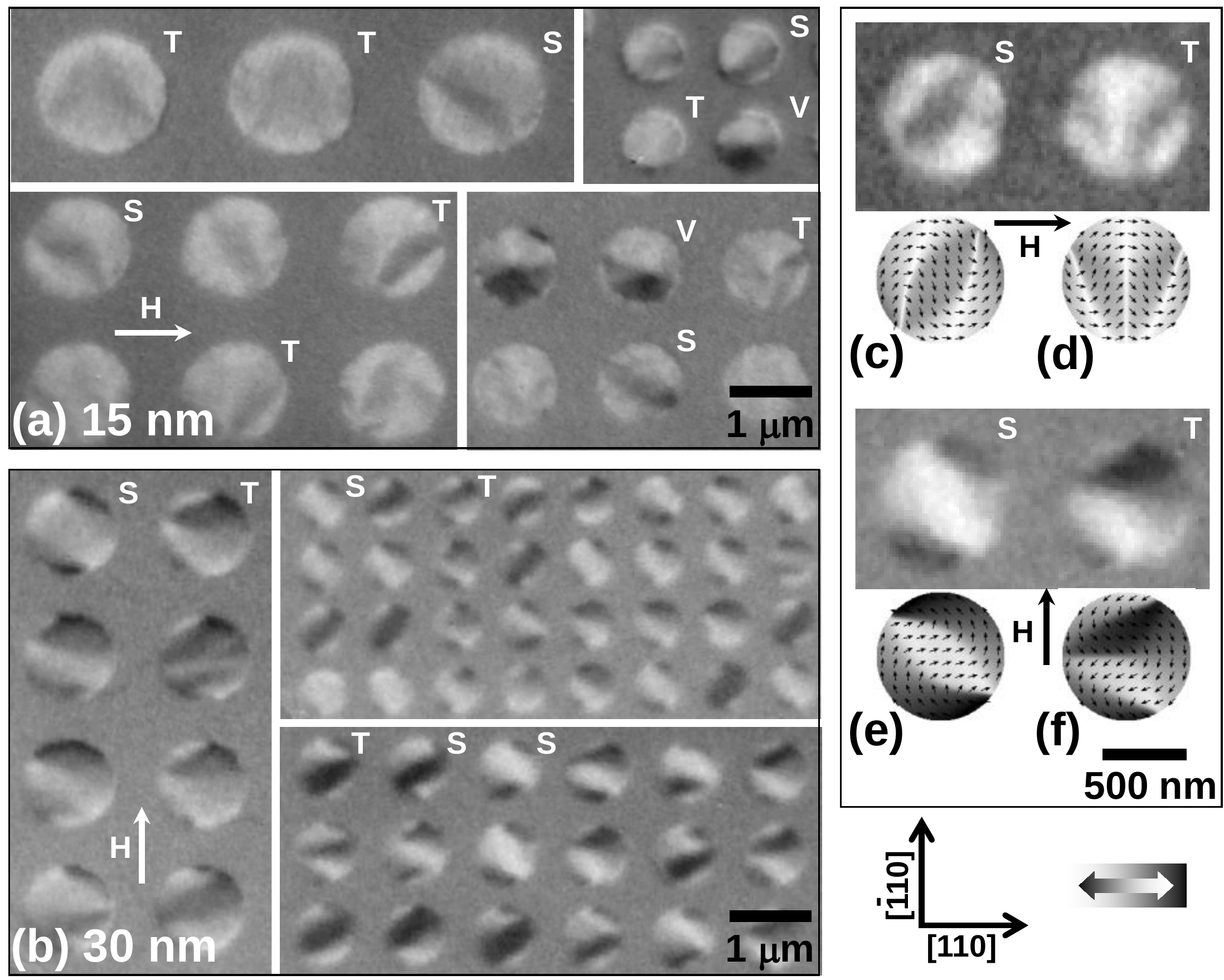}
\caption{XMCD-PEEM images of disc elements with varying size, for (a) 15 nm and (b) 30 nm \heusler\ films, with the initial magnetic field {\bf H} applied along the [110] and [$\bar{1}$10] directions, respectively (300 K). Images are representative of the states found across the larger arrays; examples of the vortex (V), triangle (T) and the S-state are shown, as labeled. (c-f) Detailed view of the triangle and S states for two 30 nm \heusler\ discs and corresponding micromagnetic simulations.}\label{fig:Heusler-Dots} 
\end{figure}

A survey of the magnetic configuration of the patterned elements after magnetic saturation (with a magnetic field of about 1 kOe) shows that high remanent states are favored, although low remanent states, such as vortex states in discs and rings, and Landau states in bar elements, are also observed. Representative magnetic states of disc elements with varying diameter, following saturation, are displayed in Fig.~\ref{fig:Heusler-Dots}. For the 15 nm thick elements, the dominant state is the so-called triangle (T) state,\cite{VKH+05} although vortex (V) states are also found, as can be observed in Fig.~\ref{fig:Heusler-Dots}(a). For the thicker elements, the triangle [Fig.~\ref{fig:Heusler-Dots}(d) and (f)] and the S-state (S) [Fig.~\ref{fig:Heusler-Dots}(c) and (e)], predicted numerically in Ref.~\onlinecite{VKH+05}, prevail. The magnetic contrast perpendicular to the average magnetization, Fig.~\ref{fig:Heusler-Dots}(b), shows that these states exhibit a significant modulation of the transverse component of the magnetization, in agreement with the results of micromagnetic simulations, Fig.~\ref{fig:Heusler-Dots}(e,f). The dominance of high remanent states points to the important role played by the magnetocrystalline anisotropy, which helps to stabilize more uniform spin configurations.\cite{VKH+05} Our results show that, for \heusler\ alloy elements below 2 $\mu$m in size, well defined magnetic states are observed, determined by the interplay between exchange energy and magnetocrystalline and shape anisotropies.

\begin{figure}[tbh]
\centering
\includegraphics[width=8.5cm]{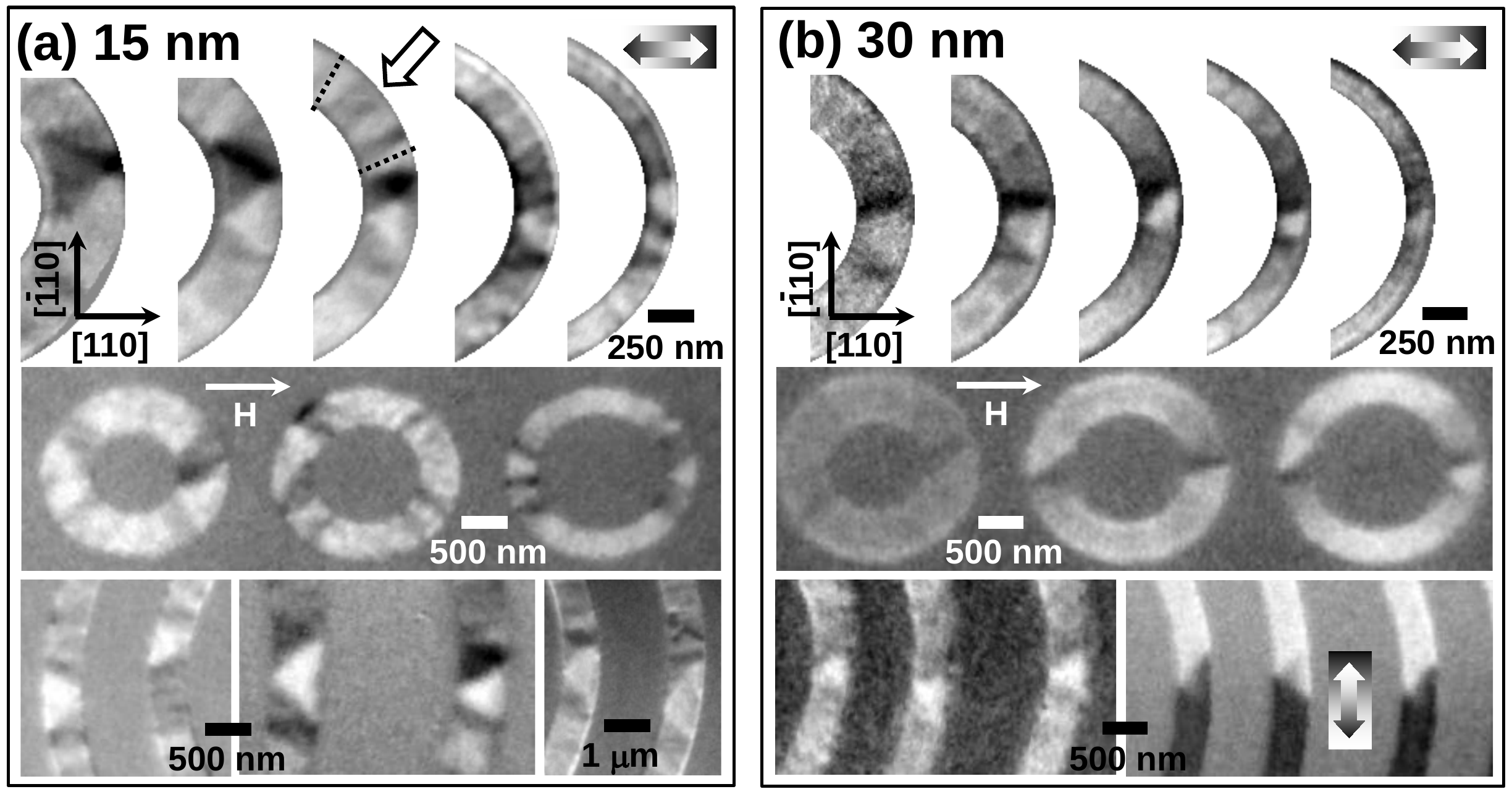}
\caption{XMCD-PEEM image of the spin configuration in rings and curved wires with varying width, initially magnetized along the [110] direction for (a) 15 nm and (b) 30 nm \heusler\ films (300 K). The arrow in (a) points to a region in the ring element showing changes in the local transverse magnetization component that resemble ripple domains in continuous films. Note the different magnetic contrast direction in the bottom right panel.}\label{fig:Heusler-SmallRings}
\end{figure}

The existence and control of domain walls is a prerequisite for domain wall based applications and experiments.\cite{Klaui08} To study domain walls in \heusler\ Heusler alloy elements, rings and zig-zag shaped wires were imaged. The magnetic configuration of the ring elements, after magnetic saturation, consists of so-called `onion states,' with domain walls that divide two quasi uniform domains in each half of the ring.\cite{RKL+01} A strong tendency for ripple domains (which are usually associated with fluctuations in the direction of the magnetic anisotropy\cite{Harte68}) is found in the wire and ring elements. The spin configuration of the onion states is characterized by the presence of local domains, as also observed in fcc Co and Fe$_3$O$_4$ ring elements due the magnetocrystalline anisotropy.\cite{KVB+03a,FHR+11} For the narrower elements we find better defined domain wall structures, as illustrated in Fig.~\ref{fig:Heusler-SmallRings}, which is explained by the stronger influence of the shape anisotropy. While transverse domain walls are dominant, we also observe vortex walls, showing that both spin configurations are stable at room temperature.\cite{Klaui08} The domain wall structure in the zig-zag wires is found to be well defined, as shown in the bottom panels of Fig.~\ref{fig:Heusler-SmallRings}. The shape anisotropy leads to head-to-head and tail-to-tail spin configurations at the wire bend, so that domain walls are formed, mostly transverse. These results show that domain walls in Heusler alloys can be generated reproducibly for elements with typical widths around 500~nm, determined mostly by the shape anisotropy. The domain wall spin configuration becomes more complicated for wider elements, which are more strongly affected by the magnetocrystalline anisotropy and magnetic ripple domains. This suggests that there is room for optimization of the film quality and/or of the patterning method to reduce local pinning.

\begin{figure}[tbh]
\centering
\includegraphics[width=5.8cm]{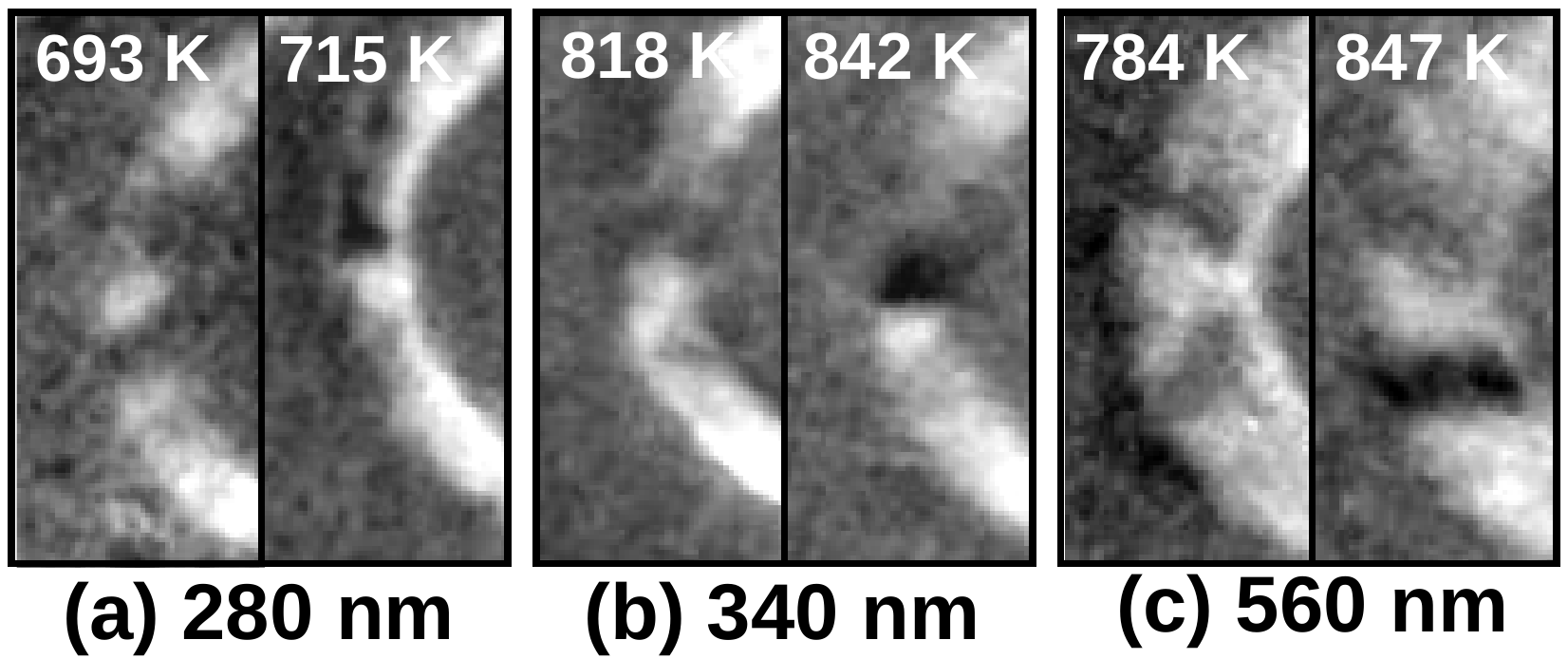}
\caption{Examples of domain wall transitions with increasing temperature (as labeled), from transverse to vortex, in 30~nm thick Heusler alloy rings with widths from 280~nm to 560~nm (a-c). For each panel, the left image shows a transverse wall at a temperature below the domain wall transition, while the image to the right shows a vortex wall at a temperature above the transition.} \label{fig:HeuslerTransition}
\end{figure}

We investigate next the thermal stability of the domain walls in ring geometries with varying widths, from 280 to 560~nm. After saturation, transverse domain walls are present in the rings, which transform progressively to vortex walls with increasing temperature (corresponding the lowest energy state, according to the results of micromagnetic simulations). Examples of transverse to vortex wall transitions with increasing temperature are presented in Fig.~\ref{fig:HeuslerTransition}, showing images before and after the transition for three ring widths taken during the heating cycle. 

\begin{figure}[b!th]
\centering
\includegraphics[width=8.0cm]{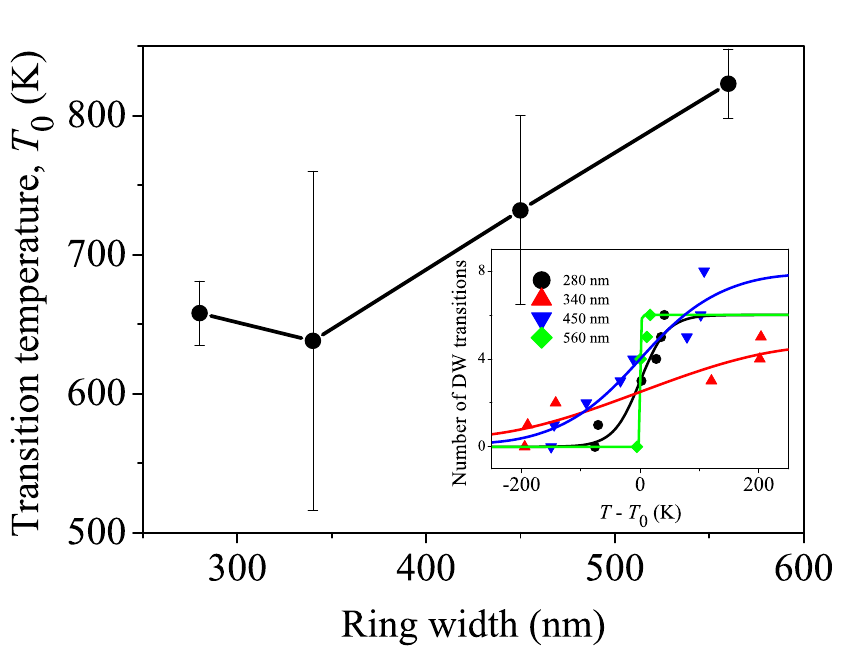}
\caption{Average transition temperature ($T_0$) as a function of ring width. Error bars correspond to the width of the transition distribution based on sampling of about six transitions per ring width, as shown in the inset (lines are Lorentzian fits to the data). For ease of display, the temperature axis in the figure inset has been offset by $T_0$.} \label{fig:HeuslerTransitionGraph} 
\end{figure}

The transition temperature is plotted as a function of ring width in Fig.~\ref{fig:HeuslerTransitionGraph}, determined from the temperature dependence of the number of domain walls that underwent a transition (out of about six to eight), shown in the inset to Fig.~\ref{fig:HeuslerTransitionGraph}. A clear dependence on the ring width is observable showing surprisingly that wider elements switch at higher temperatures, which is opposite to the expected behavior, since transverse walls tend to be favored in narrower elements.\cite{MD97,KVB+04b,NTM05,LBB+06b} At temperatures above 890 K, re-crystallization occurs due to Cr interdiffusion,\cite{TIM+06,WSS+08} leading to irreversible structural and morphological damage to the structures (not shown). The temperature of this permanent destruction depends on the element size, with narrow elements degrading first. Therefore, possible explanations for the increase in the transition temperature with increasing ring width include changes in magnetic anisotropy with temperature\cite{TIM+06,WSS+08} due to the onset of thermally activated Cr interdiffusion and crystallization processes, where smaller elements are affected at lower temperatures. Nevertheless, the heating experiments show the stability of the spin configuration up to 800~K, which marks the onset of irreversible magnetic changes in these elements. Another observation is the non-monotonic variation of the switching temperature distribution with the ring width (Fig.~\ref{fig:HeuslerTransitionGraph}, inset, and error bars in main graph), being particularly wide for the 340 nm wide rings. One possible explanation for this behavior could be a strong competition between the magnetocrystalline and shape anisotropy at this ring width, leading to a spread in the energy barrier height distribution that separates the vortex and transverse domain wall configurations.

In conclusion, the feasible control of the spin structure in patterned Heusler alloys (\heusler) is demonstrated. In confined structures with sizes of around 500~nm, the spin structure is well defined and is determined mainly by the element shape. The high spin polarization in \heusler\ elements and the resistance to thermally activated changes make this material an interesting candidate for future applications and experiments. 

This work was funded by the German Ministry for Education and Science (BMBF), project ``Multimag'' (13N9911), the Graduate School ``Material Science in Mainz'' (DFG/GSC 266), EU's 7th Framework Programme IFOX (NMP3-LA-2010 246102), MAGWIRE (FP7-ICT-2009-5 257707), the European Research Council through the Starting Independent Researcher Grant MASPIC (ERC-2007-StG 208162), the Swiss National Science Foundation, and the DFG. We also thank the team of Sensitec GmbH, Mainz for technical support during the film growth. Part of this
work was performed at the Swiss Light Source, Paul Scherrer Institut, Villigen, Switzerland.


\begin{thebibliography}{25}%
\makeatletter
\providecommand \@ifxundefined [1]{%
 \@ifx{#1\undefined}
}%
\providecommand \@ifnum [1]{%
 \ifnum #1\expandafter \@firstoftwo
 \else \expandafter \@secondoftwo
 \fi
}%
\providecommand \@ifx [1]{%
 \ifx #1\expandafter \@firstoftwo
 \else \expandafter \@secondoftwo
 \fi
}%
\providecommand \natexlab [1]{#1}%
\providecommand \enquote  [1]{``#1''}%
\providecommand \bibnamefont  [1]{#1}%
\providecommand \bibfnamefont [1]{#1}%
\providecommand \citenamefont [1]{#1}%
\providecommand \href@noop [0]{\@secondoftwo}%
\providecommand \href [0]{\begingroup \@sanitize@url \@href}%
\providecommand \@href[1]{\@@startlink{#1}\@@href}%
\providecommand \@@href[1]{\endgroup#1\@@endlink}%
\providecommand \@sanitize@url [0]{\catcode `\\12\catcode `\$12\catcode
  `\&12\catcode `\#12\catcode `\^12\catcode `\_12\catcode `\%12\relax}%
\providecommand \@@startlink[1]{}%
\providecommand \@@endlink[0]{}%
\providecommand \url  [0]{\begingroup\@sanitize@url \@url }%
\providecommand \@url [1]{\endgroup\@href {#1}{\urlprefix }}%
\providecommand \urlprefix  [0]{URL }%
\providecommand \Eprint [0]{\href }%
\providecommand \doibase [0]{http://dx.doi.org/}%
\providecommand \selectlanguage [0]{\@gobble}%
\providecommand \bibinfo  [0]{\@secondoftwo}%
\providecommand \bibfield  [0]{\@secondoftwo}%
\providecommand \translation [1]{[#1]}%
\providecommand \BibitemOpen [0]{}%
\providecommand \bibitemStop [0]{}%
\providecommand \bibitemNoStop [0]{.\EOS\space}%
\providecommand \EOS [0]{\spacefactor3000\relax}%
\providecommand \BibitemShut  [1]{\csname bibitem#1\endcsname}%
\let\auto@bib@innerbib\@empty
\bibitem [{\citenamefont {Graf}, \citenamefont {Felser},\ and\ \citenamefont
  {Parkin}(2011)}]{GFP11}%
  \BibitemOpen
  \bibfield  {author} {\bibinfo {author} {\bibfnamefont {T.}~\bibnamefont
  {Graf}}, \bibinfo {author} {\bibfnamefont {C.}~\bibnamefont {Felser}}, \ and\
  \bibinfo {author} {\bibfnamefont {S.~S.~P.}\ \bibnamefont {Parkin}},\
  }\href@noop {} {\bibfield  {journal} {\bibinfo  {journal} {Progr. Solid State
  Chem.}\ }\textbf {\bibinfo {volume} {39}},\ \bibinfo {pages} {1} (\bibinfo
  {year} {2011})}\BibitemShut {NoStop}%
\bibitem [{\citenamefont {Buschow}, \citenamefont {van Engen},\ and\
  \citenamefont {Jongebreur}(1983)}]{BEJ83}%
  \BibitemOpen
  \bibfield  {author} {\bibinfo {author} {\bibfnamefont {K.~H.~J.}\
  \bibnamefont {Buschow}}, \bibinfo {author} {\bibfnamefont {P.~G.}\
  \bibnamefont {van Engen}}, \ and\ \bibinfo {author} {\bibfnamefont
  {R.}~\bibnamefont {Jongebreur}},\ }\href@noop {} {\bibfield  {journal}
  {\bibinfo  {journal} {J. Magn. Magn. Mater.}\ }\textbf {\bibinfo {volume}
  {38}},\ \bibinfo {pages} {1} (\bibinfo {year} {1983})}\BibitemShut {NoStop}%
\bibitem [{\citenamefont {Trudel}\ \emph {et~al.}(2010)\citenamefont {Trudel},
  \citenamefont {Gaier}, \citenamefont {Hamrle},\ and\ \citenamefont
  {Hillebrands}}]{TGHH10}%
  \BibitemOpen
  \bibfield  {author} {\bibinfo {author} {\bibfnamefont {S.}~\bibnamefont
  {Trudel}}, \bibinfo {author} {\bibfnamefont {O.}~\bibnamefont {Gaier}},
  \bibinfo {author} {\bibfnamefont {J.}~\bibnamefont {Hamrle}}, \ and\ \bibinfo
  {author} {\bibfnamefont {B.}~\bibnamefont {Hillebrands}},\ }\href@noop {}
  {\bibfield  {journal} {\bibinfo  {journal} {J. Phys. D: Appl. Phys.}\
  }\textbf {\bibinfo {volume} {43}},\ \bibinfo {pages} {193001} (\bibinfo
  {year} {2010})}\BibitemShut {NoStop}%
\bibitem [{\citenamefont {Takanashi}(2010)}]{Takanashi10}%
  \BibitemOpen
  \bibfield  {author} {\bibinfo {author} {\bibfnamefont {K.}~\bibnamefont
  {Takanashi}},\ }\href@noop {} {\bibfield  {journal} {\bibinfo  {journal}
  {Jpn. J. Appl. Phys.}\ }\textbf {\bibinfo {volume} {49}},\ \bibinfo {pages}
  {110001} (\bibinfo {year} {2010})}\BibitemShut {NoStop}%
\bibitem [{\citenamefont {Ionescu}\ \emph {et~al.}(2005)\citenamefont
  {Ionescu}, \citenamefont {Vaz}, \citenamefont {Trypiniotis}, \citenamefont
  {Garc{\'\i}a-Miquel}, \citenamefont {G\"{u}rtler}, \citenamefont {Bland},
  \citenamefont {Vickers}, \citenamefont {Dalgliesh}, \citenamefont
  {Langridge}, \citenamefont {Bugoslavsky}, \citenamefont {Miyoshi},
  \citenamefont {Cohen},\ and\ \citenamefont {Ziebeck}}]{IVT+05a}%
  \BibitemOpen
  \bibfield  {author} {\bibinfo {author} {\bibfnamefont {A.}~\bibnamefont
  {Ionescu}}, \bibinfo {author} {\bibfnamefont {C.~A.~F.}\ \bibnamefont {Vaz}},
  \bibinfo {author} {\bibfnamefont {T.}~\bibnamefont {Trypiniotis}}, \bibinfo
  {author} {\bibfnamefont {H.}~\bibnamefont {Garc{\'\i}a-Miquel}}, \bibinfo
  {author} {\bibfnamefont {C.~M.}\ \bibnamefont {G\"{u}rtler}}, \bibinfo
  {author} {\bibfnamefont {J.~A.~C.}\ \bibnamefont {Bland}}, \bibinfo {author}
  {\bibfnamefont {M.~E.}\ \bibnamefont {Vickers}}, \bibinfo {author}
  {\bibfnamefont {R.~M.}\ \bibnamefont {Dalgliesh}}, \bibinfo {author}
  {\bibfnamefont {S.}~\bibnamefont {Langridge}}, \bibinfo {author}
  {\bibfnamefont {Y.}~\bibnamefont {Bugoslavsky}}, \bibinfo {author}
  {\bibfnamefont {Y.}~\bibnamefont {Miyoshi}}, \bibinfo {author} {\bibfnamefont
  {L.~F.}\ \bibnamefont {Cohen}}, \ and\ \bibinfo {author} {\bibfnamefont
  {K.~R.~A.}\ \bibnamefont {Ziebeck}},\ }\href@noop {} {\bibfield  {journal}
  {\bibinfo  {journal} {Phys. Rev. B}\ }\textbf {\bibinfo {volume} {71}},\
  \bibinfo {pages} {094401} (\bibinfo {year} {2005})}\BibitemShut {NoStop}%
\bibitem [{\citenamefont {Vaz}\ \emph {et~al.}(2010)\citenamefont {Vaz},
  \citenamefont {Hoffman}, \citenamefont {Ahn},\ and\ \citenamefont
  {Ramesh}}]{VHAR10}%
  \BibitemOpen
  \bibfield  {author} {\bibinfo {author} {\bibfnamefont {C.~A.~F.}\
  \bibnamefont {Vaz}}, \bibinfo {author} {\bibfnamefont {J.}~\bibnamefont
  {Hoffman}}, \bibinfo {author} {\bibfnamefont {C.~H.}\ \bibnamefont {Ahn}}, \
  and\ \bibinfo {author} {\bibfnamefont {R.}~\bibnamefont {Ramesh}},\
  }\href@noop {} {\bibfield  {journal} {\bibinfo  {journal} {Adv. Mater.}\
  }\textbf {\bibinfo {volume} {22}},\ \bibinfo {pages} {2900} (\bibinfo {year}
  {2010})}\BibitemShut {NoStop}%
\bibitem [{\citenamefont {Yamauchi}, \citenamefont {Sanyal},\ and\
  \citenamefont {Picozzi}(2007)}]{YSP07}%
  \BibitemOpen
  \bibfield  {author} {\bibinfo {author} {\bibfnamefont {K.}~\bibnamefont
  {Yamauchi}}, \bibinfo {author} {\bibfnamefont {B.}~\bibnamefont {Sanyal}}, \
  and\ \bibinfo {author} {\bibfnamefont {S.}~\bibnamefont {Picozzi}},\
  }\href@noop {} {\bibfield  {journal} {\bibinfo  {journal} {Appl. Phys.
  Lett.}\ }\textbf {\bibinfo {volume} {91}},\ \bibinfo {pages} {062506}
  (\bibinfo {year} {2007})}\BibitemShut {NoStop}%
\bibitem [{\citenamefont {Fecher}\ and\ \citenamefont {Felser}(2007)}]{FF07}%
  \BibitemOpen
  \bibfield  {author} {\bibinfo {author} {\bibfnamefont {G.~H.}\ \bibnamefont
  {Fecher}}\ and\ \bibinfo {author} {\bibfnamefont {C.}~\bibnamefont
  {Felser}},\ }\href@noop {} {\bibfield  {journal} {\bibinfo  {journal} {J.
  Phys. D: Appl. Phys.}\ }\textbf {\bibinfo {volume} {40}},\ \bibinfo {pages}
  {1582} (\bibinfo {year} {2007})}\BibitemShut {NoStop}%
\bibitem [{\citenamefont {Nakatani}\ \emph {et~al.}(2007)\citenamefont
  {Nakatani}, \citenamefont {Rajanikanth}, \citenamefont {Gercsi},
  \citenamefont {Takahashi}, \citenamefont {Inomata},\ and\ \citenamefont
  {Hono}}]{NRG+07}%
  \BibitemOpen
  \bibfield  {author} {\bibinfo {author} {\bibfnamefont {T.~M.}\ \bibnamefont
  {Nakatani}}, \bibinfo {author} {\bibfnamefont {A.}~\bibnamefont
  {Rajanikanth}}, \bibinfo {author} {\bibfnamefont {Z.}~\bibnamefont {Gercsi}},
  \bibinfo {author} {\bibfnamefont {Y.~K.}\ \bibnamefont {Takahashi}}, \bibinfo
  {author} {\bibfnamefont {K.}~\bibnamefont {Inomata}}, \ and\ \bibinfo
  {author} {\bibfnamefont {K.}~\bibnamefont {Hono}},\ }\href@noop {} {\bibfield
   {journal} {\bibinfo  {journal} {J. Appl. Phys.}\ }\textbf {\bibinfo {volume}
  {102}},\ \bibinfo {pages} {033916} (\bibinfo {year} {2007})}\BibitemShut
  {NoStop}%
\bibitem [{\citenamefont {Tezuka}\ \emph {et~al.}(2006)\citenamefont {Tezuka},
  \citenamefont {Ikeda}, \citenamefont {Miyazaki}, \citenamefont {Sugimoto},
  \citenamefont {Kikuchi},\ and\ \citenamefont {Inomata}}]{TIM+06}%
  \BibitemOpen
  \bibfield  {author} {\bibinfo {author} {\bibfnamefont {N.}~\bibnamefont
  {Tezuka}}, \bibinfo {author} {\bibfnamefont {N.}~\bibnamefont {Ikeda}},
  \bibinfo {author} {\bibfnamefont {A.}~\bibnamefont {Miyazaki}}, \bibinfo
  {author} {\bibfnamefont {S.}~\bibnamefont {Sugimoto}}, \bibinfo {author}
  {\bibfnamefont {M.}~\bibnamefont {Kikuchi}}, \ and\ \bibinfo {author}
  {\bibfnamefont {K.}~\bibnamefont {Inomata}},\ }\href@noop {} {\bibfield
  {journal} {\bibinfo  {journal} {Appl. Phys. Lett.}\ }\textbf {\bibinfo
  {volume} {98}},\ \bibinfo {pages} {112514} (\bibinfo {year}
  {2006})}\BibitemShut {NoStop}%
\bibitem [{\citenamefont {Shan}\ \emph {et~al.}(2009)\citenamefont {Shan},
  \citenamefont {Sukegawa}, \citenamefont {Wang}, \citenamefont {Kodzuka},
  \citenamefont {Furubayashi}, \citenamefont {Ohkubo}, \citenamefont {Mitani},
  \citenamefont {Inomata},\ and\ \citenamefont {Hono}}]{SSW+09}%
  \BibitemOpen
  \bibfield  {author} {\bibinfo {author} {\bibfnamefont {R.}~\bibnamefont
  {Shan}}, \bibinfo {author} {\bibfnamefont {H.}~\bibnamefont {Sukegawa}},
  \bibinfo {author} {\bibfnamefont {W.~H.}\ \bibnamefont {Wang}}, \bibinfo
  {author} {\bibfnamefont {M.}~\bibnamefont {Kodzuka}}, \bibinfo {author}
  {\bibfnamefont {T.}~\bibnamefont {Furubayashi}}, \bibinfo {author}
  {\bibfnamefont {T.}~\bibnamefont {Ohkubo}}, \bibinfo {author} {\bibfnamefont
  {S.}~\bibnamefont {Mitani}}, \bibinfo {author} {\bibfnamefont
  {K.}~\bibnamefont {Inomata}}, \ and\ \bibinfo {author} {\bibfnamefont
  {K.}~\bibnamefont {Hono}},\ }\href@noop {} {\bibfield  {journal} {\bibinfo
  {journal} {Phys. Rev. Lett.}\ }\textbf {\bibinfo {volume} {102}},\ \bibinfo
  {pages} {246601} (\bibinfo {year} {2009})}\BibitemShut {NoStop}%
\bibitem [{\citenamefont {Sukegawa}\ \emph {et~al.}(2010)\citenamefont
  {Sukegawa}, \citenamefont {Kasai}, \citenamefont {Furubayashi}, \citenamefont
  {Mitani},\ and\ \citenamefont {Inomata}}]{SKF+10}%
  \BibitemOpen
  \bibfield  {author} {\bibinfo {author} {\bibfnamefont {H.}~\bibnamefont
  {Sukegawa}}, \bibinfo {author} {\bibfnamefont {S.}~\bibnamefont {Kasai}},
  \bibinfo {author} {\bibfnamefont {T.}~\bibnamefont {Furubayashi}}, \bibinfo
  {author} {\bibfnamefont {S.}~\bibnamefont {Mitani}}, \ and\ \bibinfo {author}
  {\bibfnamefont {K.}~\bibnamefont {Inomata}},\ }\href@noop {} {\bibfield
  {journal} {\bibinfo  {journal} {Appl. Phys. Lett.}\ }\textbf {\bibinfo
  {volume} {96}},\ \bibinfo {pages} {042508} (\bibinfo {year}
  {2010})}\BibitemShut {NoStop}%
\bibitem [{\citenamefont {Graf}\ \emph {et~al.}(2011)\citenamefont {Graf},
  \citenamefont {Casper}, \citenamefont {Gasi}, \citenamefont {Ksenofontov},
  \citenamefont {Ruiz-Calaforra}, \citenamefont {Conca}, \citenamefont {Leven},
  \citenamefont {Jakob}, \citenamefont {Hillebrands},\ and\ \citenamefont
  {Felser}}]{GCG+11}%
  \BibitemOpen
  \bibfield  {author} {\bibinfo {author} {\bibfnamefont {T.}~\bibnamefont
  {Graf}}, \bibinfo {author} {\bibfnamefont {F.}~\bibnamefont {Casper}},
  \bibinfo {author} {\bibfnamefont {T.}~\bibnamefont {Gasi}}, \bibinfo {author}
  {\bibfnamefont {V.}~\bibnamefont {Ksenofontov}}, \bibinfo {author}
  {\bibfnamefont {A.}~\bibnamefont {Ruiz-Calaforra}}, \bibinfo {author}
  {\bibfnamefont {A.}~\bibnamefont {Conca}}, \bibinfo {author} {\bibfnamefont
  {B.}~\bibnamefont {Leven}}, \bibinfo {author} {\bibfnamefont
  {G.}~\bibnamefont {Jakob}}, \bibinfo {author} {\bibfnamefont
  {B.}~\bibnamefont {Hillebrands}}, \ and\ \bibinfo {author} {\bibfnamefont
  {C.}~\bibnamefont {Felser}},\ }\href@noop {} {} (\bibinfo {year} {2011}),\
  \bibinfo {note} {unpublished}\BibitemShut {NoStop}%
\bibitem [{OOM()}]{OOMMF}%
  \BibitemOpen
  \href@noop {} {}\bibinfo {note} {Available at
  {h}ttp://math.nist.gov/oommf}\BibitemShut {NoStop}%
\bibitem [{\citenamefont {Vaz}\ \emph {et~al.}(2005)\citenamefont {Vaz},
  \citenamefont {Kl{\"a}ui}, \citenamefont {Heyderman}, \citenamefont {David},
  \citenamefont {Nolting},\ and\ \citenamefont {Bland}}]{VKH+05}%
  \BibitemOpen
  \bibfield  {author} {\bibinfo {author} {\bibfnamefont {C.~A.~F.}\
  \bibnamefont {Vaz}}, \bibinfo {author} {\bibfnamefont {M.}~\bibnamefont
  {Kl{\"a}ui}}, \bibinfo {author} {\bibfnamefont {L.~J.}\ \bibnamefont
  {Heyderman}}, \bibinfo {author} {\bibfnamefont {C.}~\bibnamefont {David}},
  \bibinfo {author} {\bibfnamefont {F.}~\bibnamefont {Nolting}}, \ and\
  \bibinfo {author} {\bibfnamefont {J.~A.~C.}\ \bibnamefont {Bland}},\
  }\href@noop {} {\bibfield  {journal} {\bibinfo  {journal} {Phys. Rev. B}\
  }\textbf {\bibinfo {volume} {72}},\ \bibinfo {pages} {224426} (\bibinfo
  {year} {2005})}\BibitemShut {NoStop}%
\bibitem [{\citenamefont {Kl{\"a}ui}(2008)}]{Klaui08}%
  \BibitemOpen
  \bibfield  {author} {\bibinfo {author} {\bibfnamefont {M.}~\bibnamefont
  {Kl{\"a}ui}},\ }\href@noop {} {\bibfield  {journal} {\bibinfo  {journal} {J.
  Phys.: Condens. Matter}\ }\textbf {\bibinfo {volume} {20}},\ \bibinfo {pages}
  {313001} (\bibinfo {year} {2008})}\BibitemShut {NoStop}%
\bibitem [{\citenamefont {Rothman}\ \emph {et~al.}(2001)\citenamefont
  {Rothman}, \citenamefont {Kl{\"a}ui}, \citenamefont {Lopez-Diaz},
  \citenamefont {Vaz}, \citenamefont {Bleloch}, \citenamefont {Bland},
  \citenamefont {Cui},\ and\ \citenamefont {Speaks}}]{RKL+01}%
  \BibitemOpen
  \bibfield  {author} {\bibinfo {author} {\bibfnamefont {J.}~\bibnamefont
  {Rothman}}, \bibinfo {author} {\bibfnamefont {M.}~\bibnamefont {Kl{\"a}ui}},
  \bibinfo {author} {\bibfnamefont {L.}~\bibnamefont {Lopez-Diaz}}, \bibinfo
  {author} {\bibfnamefont {C.~A.~F.}\ \bibnamefont {Vaz}}, \bibinfo {author}
  {\bibfnamefont {A.}~\bibnamefont {Bleloch}}, \bibinfo {author} {\bibfnamefont
  {J.~A.~C.}\ \bibnamefont {Bland}}, \bibinfo {author} {\bibfnamefont
  {Z.}~\bibnamefont {Cui}}, \ and\ \bibinfo {author} {\bibfnamefont
  {R.}~\bibnamefont {Speaks}},\ }\href@noop {} {\bibfield  {journal} {\bibinfo
  {journal} {Phys. Rev. Lett.}\ }\textbf {\bibinfo {volume} {86}},\ \bibinfo
  {pages} {1098} (\bibinfo {year} {2001})}\BibitemShut {NoStop}%
\bibitem [{\citenamefont {Harte}(1968)}]{Harte68}%
  \BibitemOpen
  \bibfield  {author} {\bibinfo {author} {\bibfnamefont {K.~J.}\ \bibnamefont
  {Harte}},\ }\href@noop {} {\bibfield  {journal} {\bibinfo  {journal} {J.
  Appl. Phys.}\ }\textbf {\bibinfo {volume} {39}},\ \bibinfo {pages} {1503}
  (\bibinfo {year} {1968})}\BibitemShut {NoStop}%
\bibitem [{\citenamefont {Kl{\"a}ui}\ \emph {et~al.}(2003)\citenamefont
  {Kl{\"a}ui}, \citenamefont {Vaz}, \citenamefont {Bland}, \citenamefont
  {Monchesky}, \citenamefont {Unguris}, \citenamefont {Bauer}, \citenamefont
  {Cherifi}, \citenamefont {Heun}, \citenamefont {Locatelli}, \citenamefont
  {Heyderman},\ and\ \citenamefont {Cui}}]{KVB+03a}%
  \BibitemOpen
  \bibfield  {author} {\bibinfo {author} {\bibfnamefont {M.}~\bibnamefont
  {Kl{\"a}ui}}, \bibinfo {author} {\bibfnamefont {C.~A.~F.}\ \bibnamefont
  {Vaz}}, \bibinfo {author} {\bibfnamefont {J.~A.~C.}\ \bibnamefont {Bland}},
  \bibinfo {author} {\bibfnamefont {T.~L.}\ \bibnamefont {Monchesky}}, \bibinfo
  {author} {\bibfnamefont {J.}~\bibnamefont {Unguris}}, \bibinfo {author}
  {\bibfnamefont {E.}~\bibnamefont {Bauer}}, \bibinfo {author} {\bibfnamefont
  {S.}~\bibnamefont {Cherifi}}, \bibinfo {author} {\bibfnamefont
  {S.}~\bibnamefont {Heun}}, \bibinfo {author} {\bibfnamefont {A.}~\bibnamefont
  {Locatelli}}, \bibinfo {author} {\bibfnamefont {L.~J.}\ \bibnamefont
  {Heyderman}}, \ and\ \bibinfo {author} {\bibfnamefont {Z.}~\bibnamefont
  {Cui}},\ }\href@noop {} {\bibfield  {journal} {\bibinfo  {journal} {Phys.
  Rev. B}\ }\textbf {\bibinfo {volume} {68}},\ \bibinfo {pages} {134426}
  (\bibinfo {year} {2003})}\BibitemShut {NoStop}%
\bibitem [{\citenamefont {Fonin}\ \emph {et~al.}(2011)\citenamefont {Fonin},
  \citenamefont {Hartung}, \citenamefont {R{\"u}diger}, \citenamefont {Backes},
  \citenamefont {Heyderman}, \citenamefont {Nolting}, \citenamefont
  {Rodr{\'\i}guez},\ and\ \citenamefont {Kl{\"a}ui}}]{FHR+11}%
  \BibitemOpen
  \bibfield  {author} {\bibinfo {author} {\bibfnamefont {M.}~\bibnamefont
  {Fonin}}, \bibinfo {author} {\bibfnamefont {C.}~\bibnamefont {Hartung}},
  \bibinfo {author} {\bibfnamefont {U.}~\bibnamefont {R{\"u}diger}}, \bibinfo
  {author} {\bibfnamefont {D.}~\bibnamefont {Backes}}, \bibinfo {author}
  {\bibfnamefont {L.}~\bibnamefont {Heyderman}}, \bibinfo {author}
  {\bibfnamefont {F.}~\bibnamefont {Nolting}}, \bibinfo {author} {\bibfnamefont
  {A.~F.}\ \bibnamefont {Rodr{\'\i}guez}}, \ and\ \bibinfo {author}
  {\bibfnamefont {M.}~\bibnamefont {Kl{\"a}ui}},\ }\href@noop {} {\bibfield
  {journal} {\bibinfo  {journal} {J. Appl. Phys.}\ }\textbf {\bibinfo {volume}
  {109}},\ \bibinfo {pages} {07D315} (\bibinfo {year} {2011})}\BibitemShut
  {NoStop}%
\bibitem [{\citenamefont {Mc{M}ichael}\ and\ \citenamefont
  {Donahue}(1997)}]{MD97}%
  \BibitemOpen
  \bibfield  {author} {\bibinfo {author} {\bibfnamefont {R.~D.}\ \bibnamefont
  {Mc{M}ichael}}\ and\ \bibinfo {author} {\bibfnamefont {M.~J.}\ \bibnamefont
  {Donahue}},\ }\href@noop {} {\bibfield  {journal} {\bibinfo  {journal} {IEEE
  Trans. Magn.}\ }\textbf {\bibinfo {volume} {33}},\ \bibinfo {pages} {4167}
  (\bibinfo {year} {1997})}\BibitemShut {NoStop}%
\bibitem [{\citenamefont {Kl{\"a}ui}\ \emph {et~al.}(2004)\citenamefont
  {Kl{\"a}ui}, \citenamefont {Vaz}, \citenamefont {Bland}, \citenamefont
  {Heyderman}, \citenamefont {Nolting}, \citenamefont {Pavlovska},
  \citenamefont {Bauer}, \citenamefont {Cherifi}, \citenamefont {Heun},\ and\
  \citenamefont {Locatelli}}]{KVB+04b}%
  \BibitemOpen
  \bibfield  {author} {\bibinfo {author} {\bibfnamefont {M.}~\bibnamefont
  {Kl{\"a}ui}}, \bibinfo {author} {\bibfnamefont {C.~A.~F.}\ \bibnamefont
  {Vaz}}, \bibinfo {author} {\bibfnamefont {J.~A.~C.}\ \bibnamefont {Bland}},
  \bibinfo {author} {\bibfnamefont {L.~J.}\ \bibnamefont {Heyderman}}, \bibinfo
  {author} {\bibfnamefont {F.}~\bibnamefont {Nolting}}, \bibinfo {author}
  {\bibfnamefont {A.}~\bibnamefont {Pavlovska}}, \bibinfo {author}
  {\bibfnamefont {E.}~\bibnamefont {Bauer}}, \bibinfo {author} {\bibfnamefont
  {S.}~\bibnamefont {Cherifi}}, \bibinfo {author} {\bibfnamefont
  {S.}~\bibnamefont {Heun}}, \ and\ \bibinfo {author} {\bibfnamefont
  {A.}~\bibnamefont {Locatelli}},\ }\href@noop {} {\bibfield  {journal}
  {\bibinfo  {journal} {Appl. Phys. Lett.}\ }\textbf {\bibinfo {volume} {85}},\
  \bibinfo {pages} {5637} (\bibinfo {year} {2004})}\BibitemShut {NoStop}%
\bibitem [{\citenamefont {Nakatani}, \citenamefont {Thiaville},\ and\
  \citenamefont {Miltat}(2005)}]{NTM05}%
  \BibitemOpen
  \bibfield  {author} {\bibinfo {author} {\bibfnamefont {Y.}~\bibnamefont
  {Nakatani}}, \bibinfo {author} {\bibfnamefont {A.}~\bibnamefont {Thiaville}},
  \ and\ \bibinfo {author} {\bibfnamefont {J.}~\bibnamefont {Miltat}},\
  }\href@noop {} {\bibfield  {journal} {\bibinfo  {journal} {J. Magn. Magn.
  Mater.}\ }\textbf {\bibinfo {volume} {290-291}},\ \bibinfo {pages} {750}
  (\bibinfo {year} {2005})}\BibitemShut {NoStop}%
\bibitem [{\citenamefont {Laufenberg}\ \emph {et~al.}(2006)\citenamefont
  {Laufenberg}, \citenamefont {Backes}, \citenamefont {B{\"u}hrer},
  \citenamefont {Bedau}, \citenamefont {Kl{\"a}ui}, \citenamefont
  {R{\"u}diger}, \citenamefont {Vaz}, \citenamefont {Bland}, \citenamefont
  {Heyderman}, \citenamefont {Nolting}, \citenamefont {Cherifi}, \citenamefont
  {Locatelli}, \citenamefont {Belkhou}, \citenamefont {Heun},\ and\
  \citenamefont {Bauer}}]{LBB+06b}%
  \BibitemOpen
  \bibfield  {author} {\bibinfo {author} {\bibfnamefont {M.}~\bibnamefont
  {Laufenberg}}, \bibinfo {author} {\bibfnamefont {D.}~\bibnamefont {Backes}},
  \bibinfo {author} {\bibfnamefont {W.}~\bibnamefont {B{\"u}hrer}}, \bibinfo
  {author} {\bibfnamefont {D.}~\bibnamefont {Bedau}}, \bibinfo {author}
  {\bibfnamefont {M.}~\bibnamefont {Kl{\"a}ui}}, \bibinfo {author}
  {\bibfnamefont {U.}~\bibnamefont {R{\"u}diger}}, \bibinfo {author}
  {\bibfnamefont {C.~A.~F.}\ \bibnamefont {Vaz}}, \bibinfo {author}
  {\bibfnamefont {J.~A.~C.}\ \bibnamefont {Bland}}, \bibinfo {author}
  {\bibfnamefont {L.~J.}\ \bibnamefont {Heyderman}}, \bibinfo {author}
  {\bibfnamefont {F.}~\bibnamefont {Nolting}}, \bibinfo {author} {\bibfnamefont
  {S.}~\bibnamefont {Cherifi}}, \bibinfo {author} {\bibfnamefont
  {A.}~\bibnamefont {Locatelli}}, \bibinfo {author} {\bibfnamefont
  {R.}~\bibnamefont {Belkhou}}, \bibinfo {author} {\bibfnamefont
  {S.}~\bibnamefont {Heun}}, \ and\ \bibinfo {author} {\bibfnamefont
  {E.}~\bibnamefont {Bauer}},\ }\href@noop {} {\bibfield  {journal} {\bibinfo
  {journal} {Appl. Phys. Lett.}\ }\textbf {\bibinfo {volume} {88}},\ \bibinfo
  {pages} {052507} (\bibinfo {year} {2006})}\BibitemShut {NoStop}%
\bibitem [{\citenamefont {Wang}\ \emph {et~al.}(2008)\citenamefont {Wang},
  \citenamefont {Sukegawa}, \citenamefont {Shan}, \citenamefont {Furubayashi},\
  and\ \citenamefont {Inomata}}]{WSS+08}%
  \BibitemOpen
  \bibfield  {author} {\bibinfo {author} {\bibfnamefont {W.}~\bibnamefont
  {Wang}}, \bibinfo {author} {\bibfnamefont {H.}~\bibnamefont {Sukegawa}},
  \bibinfo {author} {\bibfnamefont {R.}~\bibnamefont {Shan}}, \bibinfo {author}
  {\bibfnamefont {T.}~\bibnamefont {Furubayashi}}, \ and\ \bibinfo {author}
  {\bibfnamefont {K.}~\bibnamefont {Inomata}},\ }\href@noop {} {\bibfield
  {journal} {\bibinfo  {journal} {Appl. Phys. Lett.}\ }\textbf {\bibinfo
  {volume} {92}},\ \bibinfo {pages} {221912} (\bibinfo {year}
  {2008})}\BibitemShut {NoStop}%
\end{thebibliography}

%

\end {document}